\documentclass[conference,9pt]{IEEEtran}
\IEEEoverridecommandlockouts

\usepackage{svg}
\usepackage{array}
\usepackage{tabularx}
\usepackage{cite}
\usepackage{amsmath,amssymb,amsfonts}
\usepackage{algorithmic}
\usepackage{graphicx}
\usepackage{textcomp}
\usepackage{xcolor}
\usepackage{arydshln}
\usepackage{boldline, makecell, hhline, pifont}

\def\BibTeX{{\rm B\kern-.05em{\sc i\kern-.025em b}\kern-.08em
    T\kern-.1667em\lower.7ex\hbox{E}\kern-.125emX}}
\begin{document}
\title{FlowSE: Flow Matching-based Speech Enhancement\\
\thanks{This research was supported by the MSIT (Ministry of Science and ICT), Korea, under the ITRC (Information Technology Research Center) support program (IITP-2024-2021-0-01835) and Artificial Intelligence Graduate School Program (No.2019-0-01842) supervised by the IITP (Institute of Information Communications Technology Planning Evaluation).

        Source codes are available online at: https://github.com/seongq/flowmse
        }
}



\author{
\IEEEauthorblockN{Seonggyu Lee}
\IEEEauthorblockA{
\textit{Gwangju Institute of Science and Technology} \\
Gwangju, Korea \\
lsqjin2022@gm.gist.ac.kr}
\\
\IEEEauthorblockN{Sangwook Han}
\IEEEauthorblockA{
\textit{Gwangju Institute of Science and Technology} \\
Gwangju, Korea \\
swhan9873@gm.gist.ac.kr}
\and
\IEEEauthorblockN{Sein Cheong}
\IEEEauthorblockA{
\textit{Gwangju Institute of Science and Technology} \\
Gwangju, Korea \\
seiinjung@gm.gist.ac.kr}
\\
\IEEEauthorblockN{Jong Won Shin}
\IEEEauthorblockA{
\textit{Gwangju Institute of Science and Technology} \\
Gwangju, Korea \\
jwshin@gist.ac.kr}
}
\maketitle

\begin{abstract}
Diffusion probabilistic models have shown impressive performance for speech enhancement, but they typically require 25 to 60 function evaluations in the inference phase, resulting in heavy computational complexity. 
Recently, a fine-tuning method was proposed to correct the reverse process, which significantly lowered the number of function evaluations (NFE). 
Flow matching is a method to train continuous normalizing flows which model probability paths from known distributions to unknown distributions including those described by diffusion processes.
In this paper, we propose a speech enhancement based on conditional flow matching.
The proposed method achieved the performance comparable to those for the diffusion-based speech enhancement with the NFE of 60 when the NFE was 5, and showed similar performance with the diffusion model correcting the reverse process at the same NFE from 1 to 5 without additional fine tuning procedure. 
We also have shown that the corresponding diffusion model derived from the conditional probability path with a modified optimal transport conditional vector field demonstrated similar performances with the NFE of 5 without any fine-tuning procedure. 

\end{abstract}


\section{Introduction}
\label{sec:intro}
The objective of speech enhancement (SE) is to recover the clean speech signals from those corrupted by environmental noises \cite{loizou2007speech, hendriks2022dft}.
Classical approaches often utilize statistical properties of the clean speech signals and the environmental noises \cite{gerkmann2018spectral, minseungenhancement, seinenhancement, ephraimmmsestsa}, but most of the recent studies utilize deep neural networks (DNNs) to estimate clean speech signals or masks to be multiplied to noisy signals or features \cite{hyungchanenhancement , hu2020dccrn ,enhancementwangtan, enhancementwangtandilated}. 
Recently, generative approaches for SE learning the underlying distribution of clean speech signals have been proposed \cite{enhancementgan,enhancementflow, enhancementvae,sgmsep,refgerkmannenhancement1 ,lemercier2023storm, Lay2023bbed, diffheun,vpidm, lay2024singlesgmsecrp}.
Among them, score-based generative models or diffusion models with stochastic differential equations (SDEs) have shown impressive results for SE \cite{sgmsep,refgerkmannenhancement1 ,lemercier2023storm, Lay2023bbed, diffheun,vpidm, lay2024singlesgmsecrp}.
An estimate of the clean speech signal is obtained by solving the reverse SDEs of diffusion models, which requires the estimation of the scores using DNNs repeatedly.
The number of function evaluations (NFEs), \emph{i.e.,} the number of running a DNN model to evaluate scores in the inference phase, were more than 25 in \cite{sgmsep,refgerkmannenhancement1 ,lemercier2023storm, Lay2023bbed, diffheun,vpidm}, which may limit the applications for the diffusion model-based SE.
In \cite{lay2024singlesgmsecrp}, a fine-tuning method for the score network was proposed, which reduced the NFE to 5 with similar performance to the method in \cite{Lay2023bbed} with the NFE of 60 by correcting the reverse process (CRP).

\setlength{\abovecaptionskip}{0pt}   
\setlength{\belowcaptionskip}{0pt}   

\begin{figure}[!t]
    \centering
    \includegraphics[width=0.485\textwidth]{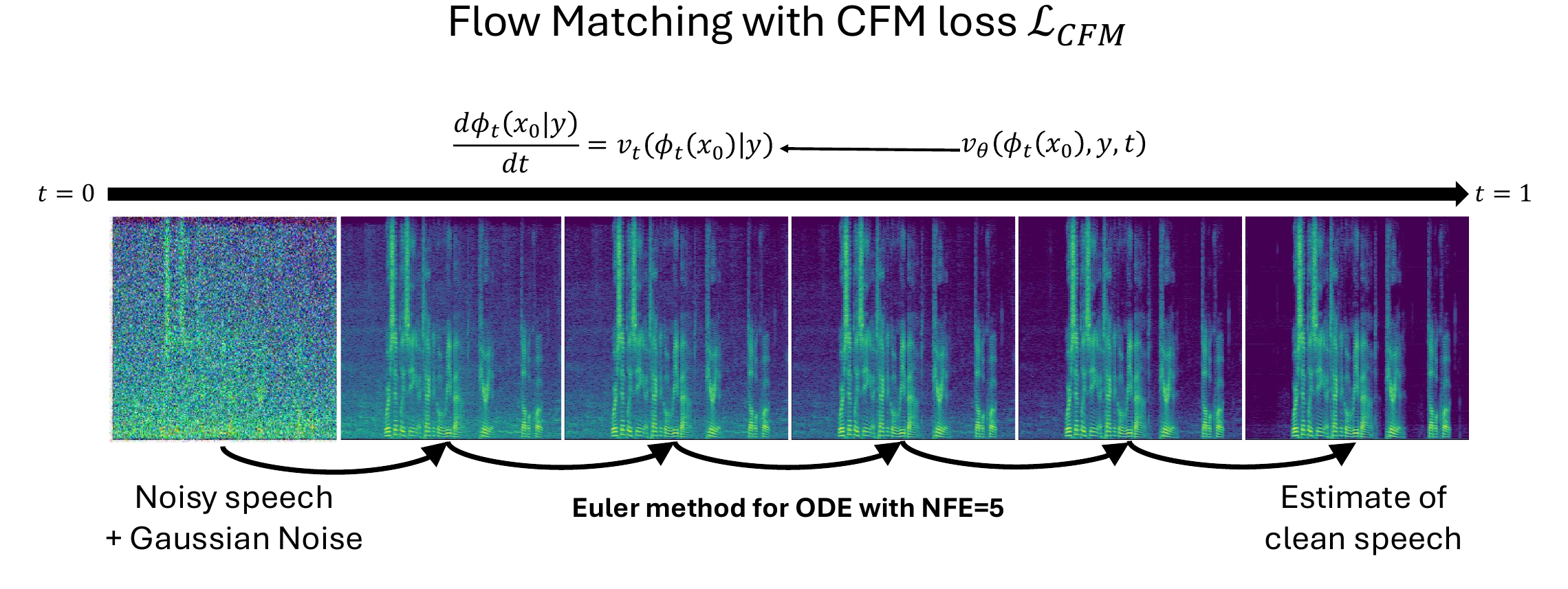} 
    \caption{Illustration of an exemplary trajectory of an ODE trained with CFM loss, which 
     transforms noisy speech corrupted by Gaussian noise into clean speech using the Euler method with the NFE of 5.}
    
    \label{figure:flowmatching}
      \vspace{-0.7cm} 
\end{figure}
As a potential alternative to the diffusion models, flow matching (FM) to train continuous normalizing flows (CNFs) has been proposed \cite{lipmanflow,tong2024improving}. FM has been utilized in applications such as generative models to downstream tasks in speech processing \cite{speechflow} and audio-visual speech enhancement \cite{flowavse}.
A flow is an invertible mapping which gradually transforms a random vector following a simple distribution into another random vector with a complex distribution, and a CNF describes a flow with an oridinary differential equation (ODE) \cite{chen2018neural}.
The FM tries to estimate the time-dependent vector field, which is the time derivative of the flow, but the ground-truth value for the vector field is often intractable. 
In \cite{lipmanflow}, the conditional flow matching (CFM) which estimates conditional vector fields given the training sample following the complex distribution was proposed, which is proven to be equivalent to the FM. 
As an example, optimal transport (OT) conditional vector field was suggested, providing faster sampling and better performance than previous diffusion models \cite{lipmanflow}.
Furthermore, it was shown that existing diffusion models can be trained with FM, leading faster sampling compared to training with score matching (SM) used for the diffusion models \cite{lipmanflow}. 

In this paper, we propose FlowSE: flow matching-based speech enhancement as an alternative to previous diffusion-based speech enhancement models, which is illustrated in Fig. \ref{figure:flowmatching}.
In the FlowSE, the noisy speech is used as a conditioning variable along with the clean speech for the conditional flow and the conditional vector field, and the OT conditional vector field is modified to incorporate the noisy speech.
Our experimental results demonstrate that the performance of FlowSE with the NFE of 5 is comparable to that of the previous diffusion model with the NFE of 60, and the fine-tuned CRP model. 
We also present an SDE to explain how FlowSE can be considered as a diffusion model and train this diffusion model using SM, achieving similar performance to FlowSE trained with CFM when the NFE was 5 without the fine-tuning procedure in CRP. 
\section{Background}
\subsection{Speech Enhancement using Diffusion Models}
SE is the process of estimating a clean speech $s$ from a noisy speech $y=s+n$, suppressing an additive environmental noise $n$. 
In the diffusion model-based SE \cite{refgerkmannenhancement1,diffheun,vpidm,sgmsep,lemercier2023storm, Lay2023bbed, lay2024singlesgmsecrp}, a diffusion process which gradually transforms a clean speech $x_0$ into a mixture of the noisy speech and Gaussian noise $x_T$ is defined as a following forward SDE: 
\begin{equation}
\label{eq:forwardsde}
    dx_t = f(x_t,y,t) dt + g(t)dw_t,
\end{equation}
where $t\in [0,T]$, $w_t$ is a Brownian motion, $f$ and $g$ are called the drift and the diffusion coefficients, respectively.
The reverse process to transform $\tilde{x}_T$ back to $\tilde{x}_0$ can be described by a reverse SDE \cite{song2021scorebased}:
\begin{equation}
    \label{eq:reversesde}
    d\tilde{x}_t = \left[f(\tilde{x}_t,y,t)-g(t)^2 \nabla_{\tilde{x}_t} \log p_t(\tilde{x}_t | y) \right]dt + g(t) d \bar{w}_t,
\end{equation}
where $\nabla_{x_t} \log p_t({x}_t|y)$, called a score function, is the gradient of log of the probability density function (pdf) of ${x}_t$ given $y$, and $\bar{w}_t$ is a reverse Brownian motion.
As the $f$ and $g$ are pre-defined functions, the estimate of the clean speech can be obtained by solving the reverse SDE ($\ref{eq:reversesde}$) from $t=T$ to $0$ with a starting point $\tilde{x}_T$ sampled from 
$\mathcal{N}(y, \sigma_T^2\mathbf{I})$, once the score function $\nabla_{x_t} \log p_t(x_t|y)$ is estimated.
The score function is approximated by a neural network called a score network $s_\theta(x_t, y, t)$ parametrized by $\theta$, which is trained with 
the denoising score matching (DSM) loss \cite{song2021scorebased} $\mathcal{L}_{DSM}(\theta)$ defined as
\begin{align}
     &\mathcal{L}_{DSM}(\theta):= \nonumber
     \\&\mathbf{E}_{t,(x_0,y),  p_t(x_t \mid x_0,y)} \lVert s_{\theta}(x_t,y,t)- \nabla_{x_t} \log p_t(x_t | x_0,y)\rVert^2,
    \label{eq:dsmloss}
\end{align}
where $t$ follows a uniform distribution between $0$ and $T$, $\mathcal{U}[0, T]$, and 
$x_t$ is sampled from a perturbation kernel $p_t(x_t | x_0, y) =\mathcal{N}\left(\mu_t(x_0,y),\sigma_t^2 \mathbf{I}\right)$, in which $\mu_t$ and $\sigma_t^2$ depend on $f$ and $g$.

\subsection{Brownian Bridge with Exponential Diffusion Coefficient (BBED)}
In \cite{Lay2023bbed}, a diffusion model with the SDE called Brownian Bridge with Exponential Diffusion Coefficient (BBED) is proposed, which has shown the best performance on the WSJ0-CHiME3 dataset 
to the best our knowledge.
The drift and diffusion coefficients for the forward BBED SDE are given by 
\begin{equation}
\label{eq:driftbbed}
    f(x_t,y,t) = \frac{y-x_t}{1-t},
\end{equation}
\begin{equation}
\label{eq:diffbbed}
    g(t) = ck^t, \text{ where } c,k>0,
\end{equation}
for $0\leq t \leq T\leq 1$. 
It is reported that the best result of BBED was achieved with the NFE of 60 when the time point at which the reverse SDE started, $t_{rsp}$, equals to $T=0.999$,  but it is also shown that a slightly lower performance was achieved with the NFE of 30 when $t_{rsp}$ was $0.5$ \cite{Lay2023bbed}. 

\subsection{Correcting the Reverse Process (CRP)}
To reduce the NFE of diffusion model BBED for SE, a fine-tuning method called CRP \cite{lay2024singlesgmsecrp} was proposed to correct the error in discretizing reverse SDE too coarsely by actually running the reverse SDE for each training sample and minimizing the estimation error \cite{lay2024singlesgmsecrp}.
Let $s_{\theta,BBED}(x_t,y,t)$ be the pre-trained score model of BBED, $n_{rev}$ be a fixed positive integer denoting the number of time points used in the discretized reverse process, and $t_\delta\approx 0$ be a sufficiently small positive real number.
Let $t_{n_{rev}}=t_{rsp} > t_{n_{rev}-1} > \cdots > t_{1} = t_\delta> t_0=0$ and $\Delta t(i) = t_{i-1}-t_{i}$.
The estimate of the clean speech, $\tilde{x}_0$, can be obtained by solving the discretized reverse SDE using the Euler-Maruyama (EuM) method \cite{song2021scorebased} starting from $\tilde{x}_{t_{n_{rev}}}$ sampled from $\mathcal{N}(y, \sigma_{t_{n_{rev}}}^2 \mathbf{I})$, which is given by 
\begin{align}
    \tilde{x}_{t_{i-1}} &= \left[f(\tilde{x}_{t_{i}}, y, t_i) - g(t_{i})^2 s_{\theta,\text{BBED}}(\tilde{x}_{t_i},y,t_i) \right]\Delta t(i) \label{eq:EuM}\\
    & \quad + g(t_{i}) \sqrt{-\Delta t (i)} \epsilon_{t_{i}}+\tilde{x}_{t_i}, \nonumber 
\end{align}
where $\epsilon_{t_{i}}$ is sampled from $\mathcal{N}(\mathbf{0}, \mathbf{I})$. 
The parameters of $s_{\theta,BBED}(\cdot, \cdot, t_1)$ is fine-tuned with the CRP loss, $\mathcal{L}_{CRP}(\theta)$, which is the mean square error between $\tilde{x}_0$ and $x_0$, i.e., 
\begin{equation}
\label{eq:crploss}
    \mathcal{L}_{CRP}(\theta) = \mathbf{E}_{y, (x_0,y), \tilde{x}_{0} | (x_0,y)} \lVert \tilde{x}_0 - x_0 \rVert ^2.
\end{equation}
As reported in \cite{lay2024singlesgmsecrp}, it was shown that CRP with $n_{rev}=5$ evaluated with the NFE of 5 achieved a nearly equivalent performance to BBED with the NFE of 60, with the proposed fine-tuning procedure.


\subsection{Flow matching}

FM \cite{lipmanflow, tong2024improving} is a method to train CNFs \cite{chen2018neural}.
CNFs are continuous time versions of normalizing flows \cite{pmlr-v37-rezende15}, which models an invertible mapping from a simple space with a known distribution $p(x_0)$, \emph{e.g.}, $\mathcal{N}\left(\mathbf{0}, \mathbf{I}\right)$, to another space distributed by an unknown distribution $q(x_1)$, when only samples from $q(x_1)$ are accessible. 
The invertible mapping $\phi_t : [0,1] \times \mathbb{R}^d \to \mathbb{R}^d$, called a flow, is defined by a time-dependent vector field $v_t:[0,1] \times \mathbb{R}^d \to \mathbb{R}^d$ with an ODE with an initial condition at $t=0$:
\begin{equation}
\label{eq:odecnf}
    \frac{d}{dt} \phi_t(x_0) = v_t(\phi_t(x_0)), ~~\phi_0(x_0) = x_0
\end{equation}
where $d$ is a dimension of a data sample.
It is noted that the pdf of $x_t=\phi_t(x_0)$, $p_t(x_t)$, can be represented as 
\begin{equation}
    p_t(x_t)= p_0 (\phi_t^{-1} (x_t)) \det \left[ \frac{\partial \phi_t^{-1}(x_t)}{\partial x_t} \right].
\end{equation}
$p_t(x_t)$ for $t\in[0,1]$, $p_t: [0,1] \times \mathbb{R}^d \to [0, \infty)$, is called a probability density path or a probability path generated by a flow $\phi_t$ or a time-dependent vector field $v_t$. 

Training CNFs aims to estimate a flow $\phi_t$, or equivalently a time-dependent vector field $v_t$ such that 
$p_0(x_0):=p(x_0)$ and 
$p_1(x_1)\approx q(x_1)$.
The vector field $v_t(x)$ generating the probability density path $p_t$ can be learned by a neural network $v_{\theta}(x,t)$ \cite{chen2018neural}. 
If a target probability density path $p_t$ and a vector field $v_t$ generating $p_t$ are given, the target vector field can be learned by optimizing the FM loss $\mathcal{L}_{FM}(\theta)$, which is defined by
\begin{equation}
    \label{eq:fmobjective} \mathcal{L}_{FM}(\theta) := \mathbf{E}_{t, p_t(x_t)} \lVert v_{\theta}(x_t,t)-v_t(x_t) \rVert^2,
\end{equation}
where $t\sim \mathcal{U}[0,1]$.
However, the FM loss can not be used in practice, because the vector field $v_t$ and the target probability path $p_t$ are often intractable. 
In \cite{lipmanflow}, a CFM loss is proposed, in which mathematically tractable $p_t(x_t|x_1)$ and $v_t(x_t|x_1)$ are considered instead of $p_t(x_t)$ and $v_t(x_t)$. The conditional flow $\phi_t(x_0|x_1)$ and the conditional vector field $v_t(x_t|x_1)$ generate the conditional probability path $p_t(x_t|x_1)$ and follow the same ODE in \eqref{eq:odecnf}. 
The CFM loss given by 
\begin{equation}
    \label{eq:cfmobjective}
    \mathcal{L}_{CFM}(\theta) := \mathbf{E}_{t,x_1, p_t(x_t\mid x_1)} \lVert v_{\theta}(x_t,t) - v_t(x_t | x_1)\rVert^2.
\end{equation}
satisfies $\nabla_{\theta}\mathcal{L}_{FM}(\theta) = \nabla_{\theta}\mathcal{L}_{CFM}(\theta)$ under mild regularity conditions \cite{lipmanflow}. 
 For a Gaussian conditional probability path $p_t(x_t|x_1)=\mathcal{N}(\mu_t(x_1), \sigma_t^2(x_1) \mathbf{I})$ in which $\mu_0$ and $\sigma_0$ do not depend on $x_1$, \emph{i.e.,} $p_0(x_0|x_1)=p_0(x_0)=p(x_0)$, the conditional flow and the conditional vector field generating $p_t(x_t|x_1)$ have closed forms:
\begin{equation}
    \phi_t(x_0| x_1) = \frac{\sigma_t (x_1)}{\sigma_0 }\left(x_0-\mu_0\right)+\mu_t(x_1) \label{eq:condflow},
\end{equation}
\begin{equation}
    v_t(x_t|x_1) = \frac{\sigma_t^\prime (x_1)}{\sigma_t (x_1)}(x_t-\mu_t(x_1))+\mu_t^\prime(x_1),
\end{equation}
where $\sigma_t^\prime = \frac{d}{dt} \sigma_t, \mu_t^{\prime} = 
\frac{d}{dt} \mu_t$. 
Since $x_t= \phi_t(x_0|x_1)$, the CFM loss in (\ref{eq:cfmobjective}) can be alternatively written as 
\begin{equation}
        \mathcal{L}_{CFM}(\theta):=\mathbf{E}_{t , x_1  , p(x_0)} \lVert v_{\theta}(\phi_t(x_0|x_1),t) - v_t(\phi_t(x_0|x_1)|x_1) \rVert^2.
\end{equation}
It is noted that $\mu_t$ and $\sigma_t$ should be pre-determined to define the Gaussian conditional path $p_t(x_t|x_1) = \mathcal{N}\left( \mu_t(x_1), \sigma_t^2(x_1) \mathbf{I}\right)$. 
As a useful example, the mean and the standard deviation corresponding to OT conditional vector field were introduced in \cite{lipmanflow}:
\begin{equation}
    \mu_t(x_1) = tx_1,~~
    \sigma_t = 1- (1-\sigma )t.
    \label{eq:OT}
\end{equation}

\section{Method}
\subsection{FlowSE: Flow Matching-based Speech Enhancement}
To adopt the flow matching for speech enhancement in which noisy speech $y$ is given, the ODE of the flow in \eqref{eq:odecnf} is modified to incorporate $y$ in a similar way to the SDE in \eqref{eq:forwardsde}:
\begin{equation}
\label{eq:flowmatchingodeSE}
    \frac{d\phi_t(x_0|y)}{dt} = v_t(\phi_t(x_0|y)|y), \phi_0(x_0|y) = x_0
\end{equation}
in which the flow $\phi_t(x_0|y)$ and vector field $v_t(x_t|y)$ conditioned on $y$ generate the probability path $p_t(x_t|y)$ from $p_0(x_0|y) = \mathcal{N}( y, \sigma^2 \mathbf{I})$ to $p_1(x_1|y)\approx q(x_1|y)$, where $q(s|y)$ is the distribution of clean speech given the noisy speech and $\sigma$ is a hyperparameter satisfying $\sigma \geq 0 $. 
Then, we estimate the vector field $v_t(x_t|y)$ by a neural network $v_{\theta} (x_t,y,t)$ using the CFM loss so that $x_1$ can be found by solving the ODE. The CFM loss for SE is given by 
\begin{align}
   & \mathcal{L}_{CFM}(\theta) := \nonumber \\
   & \mathbf{E}_{t, (x_1,y), p(x_0 |y)} \lVert v_{\theta} (\phi_t(x_0|x_1,y),y,t) - v_t(\phi_t(x_0|x_1,y)|x_1,y)  \rVert^2.
    \label{eq:cfmenhancement}
\end{align}

As in \cite{lipmanflow}, we let $p_t(x_t | x_1,y)$ be a Gaussian conditional probability path. Then, the conditional flow $\phi_t(x_0|x_1,y)$ and the conditional target vector field $v_t(x_t |x_1,y)$ becomes
\begin{equation}
    \phi_t(x_0| x_1, y) = \frac{\sigma_t }{\sigma_0 }\left(x_0-\mu_0(x_1,y)\right)+\mu_t(x_1,y) \label{eq:condflowSE},
\end{equation}
\begin{equation}
    v_t(x_t|x_1,y) = \frac{\sigma_t^\prime }{\sigma_t }(x_t-\mu_t(x_1,y))+\mu_t^\prime(x_1,y), 
\end{equation}
in which the mean $\mu_t(x_1,y)$ and variance $\sigma_t^2$ for $p_t(x_t | x_1,y)= \mathcal{N}\left( \mu_t(x_1, y), \sigma_t^2 \mathbf{I}\right)$ are configured in a similar way to the OT conditional vector field in \eqref{eq:OT} as 
\begin{equation}
    \mu_t(x_1,y) = tx_1+(1-t)y, \label{eq:mean} 
\end{equation}
\begin{equation}
    \sigma_t=   (1-t)\sigma. \label{eq:standarddev}
\end{equation}
In (\ref{eq:mean}), the mean moves linearly from the noisy speech $y$ to the clean speech $x_1$ from $t=0$ to $1$, while $\mu_t$ in (\ref{eq:OT}) starts at $0$ and goes linearly to $x_1$. 
The standard deviation in (\ref{eq:standarddev}) decreases linearly from $\sigma$ at $t=0$ to $0$ at $t=1$.
In contrast, the standard deviation $\sigma_t$ in (\ref{eq:OT}) decreases from 1 at $t=0$ to a small positive value $\sigma<1$. In the training procedure, $t$ is sampled from $\mathcal{U}[0,1-t_\delta]$ for each training sample pair $(x_1, y)$, and then $x_t$ is directly sampled from $p_t(x_t|x_1,y)$, which is used to construct the training target in \eqref{eq:cfmenhancement}.

In the inference phase, the ODE in (\ref{eq:flowmatchingodeSE}) is solved numerically using Euler method with trained $v_{\theta}(x_t,y,t)$ starting at $x_0$ sampled from $p_0(x_0|y)$. 
$N$ time points are chosen on $[0,1]$ and denoted as $0=t_0<t_1<\cdots<t_{N-1}=1-t_\delta<t_N=1$. 
Starting from the initial point $x_0$ from $p_0(x_0|y)$, the clean speech estimate $x_{t_{N}}$ is generated by applying the following equation repeatedly:
\begin{equation}
    x_{t_{i}}= v_{\theta}(x_{t_{i-1}},y, t_{i-1} ) \Delta t(i)+x_{t_{i-1}},\label{eq:eulerforSE}
\end{equation}
where $\Delta t(i) = t_i - t_{i-1}$.

\subsection{FlowSE as a Diffusion model}
The proposed FlowSE with a Gaussian conditional probability path can also be described as a diffusion model with an SDE. We can find the drift coefficient $f$ and the diffusion coefficient $g$ in (\ref{eq:forwardsde}) which make the perturbation kernel ${p}_t(x_t | x_0,y)$ match the conditional probability path ${p}_t(x_t | x_1,y)$ in the FM with reversed time, \emph{i.e.}, ${p}_t(x_t | x_0,y)=\mathcal{N}( \mu_{1-t} (x_0,y), \sigma_{1-t}^2\mathbf{I} )$. By utilizing the ODEs that the mean and the covariance for the perturbation kernel ${p}_t(x_t | x_0,y)$ of a diffusion model satisfy, which is (5.50), (5.53) in \cite{appliedsde}, the drift and diffusion coefficients for the equivalent diffusion model are given by 
\begin{equation}
    f(x_t,y,t)  = \frac{y-x_t}{1-t},  ~~g(t) =\sqrt{ \frac{2t\sigma^2}{1-t}}.
    \label{eq:driftdiffusion}
\end{equation}
It is noted that the drift coefficient $f$ is the same as the $f$ in the BBED SDE given in \eqref{eq:driftbbed}, while the diffusion coefficient is different. 
One of the method to find $x_0$ from the diffusion model in \eqref{eq:forwardsde} is to solve the probability flow ODE \cite{song2021scorebased} given by 
\begin{equation}
    \frac{dx_t}{dt} = \left[f(x_t,y,t)- \frac{1}{2} g(t)^2 \nabla_{x_t} \log p_t(x_t | y) \right],
    \label{eq:pfode}
\end{equation}
which has the form of the ODE in \eqref{eq:flowmatchingodeSE} \cite{lipmanflow, tong2024improving}. It can be solved using the Euler method, moving from $t_N \in (0,1)$ to $t_0=0$, with $s_\theta(x_t,y,t)$ trained with the DSM loss, instead of $\nabla_{x_t} \log p_t (x_t |y)$. 
\begin{table*}[!t]
\caption{Speech enhancement performances for proposed and compared methods on the WSJ0-CHiME3 and VB-DMD datasets.}
\centering
\renewcommand{\arraystretch}{1.2} 
\setlength{\tabcolsep}{4pt} 

\scriptsize

\label{maintable}
\begin{tabularx}{\textwidth}{c>{\centering\arraybackslash}X>{\centering\arraybackslash}X>{\centering\arraybackslash}X>{\centering\arraybackslash}X>{\centering\arraybackslash}X>{\centering\arraybackslash}X>{\centering\arraybackslash}X>{\centering\arraybackslash}X>{\centering\arraybackslash}X>{\centering\arraybackslash}X>{\centering\arraybackslash}X>{\centering\arraybackslash}X}
\hlineB{2}
\multicolumn{12}{c}{Tranined and Tested on WSJ0-CHiME3} \\ 
\hlineB{2}
Method & NFE  & PESQ & DNSMOS & WVMOS & ESTOI & SIG & BAK & OVRL & SI-SDR & SI-SIR & SI-SAR \\
\hlineB{2}
BBED  & 60   & 3.00$\pm$0.05&\textbf{4.04$\pm$0.01}&3.93$\pm$0.03&0.93$\pm$0.00&3.59$\pm$0.01&4.18$\pm$0.00&3.35$\pm$0.01&18.98$\pm$0.35&31.48$\pm$0.39&19.31$\pm$0.36
 \\
BBED  & 5   & 2.90$\pm$0.04&3.86$\pm$0.02&3.75$\pm$0.03&0.92$\pm$0.00&3.53$\pm$0.01&4.05$\pm$0.01&3.23$\pm$0.01&18.33$\pm$0.35&26.82$\pm$0.38&19.04$\pm$0.36
\\
CRP  & 5 & 3.02$\pm$0.05&4.01$\pm$0.01&3.90$\pm$0.03&\textbf{0.94$\pm$0.00}&3.55$\pm$0.01&4.17$\pm$0.00&3.31$\pm$0.01&\textbf{19.71$\pm$0.34}&30.77$\pm$0.39&\textbf{20.14$\pm$0.34}
 \\
Ours & 5 & \textbf{3.03$\pm$0.04} &\textbf{4.04$\pm$0.01}
&\textbf{3.96$\pm$0.03}
&0.93$\pm$0.00&\textbf{3.60$\pm$0.01}&\textbf{4.20$\pm$0.00}&\textbf{3.37$\pm$0.01}&18.95$\pm$0.33&\textbf{32.43$\pm$0.41}&19.21$\pm$0.34\\

\hlineB{2}
\multicolumn{12}{c}{ Tranined and Tested on VB-DMD } \\ 
\hlineB{2}
BBED & 60  & 3.09$\pm$0.05&3.57$\pm$0.02&4.29$\pm$0.02&\textbf{0.88$\pm$0.01}&3.48$\pm$0.01&4.04$\pm$0.01&3.2$\pm$0.01&18.75$\pm$0.23&30.11$\pm$0.41&19.43$\pm$0.24
 \\
BBED & 5 & 2.43$\pm$0.04&3.3$\pm$0.02&3.93$\pm$0.04&0.86$\pm$0.01&3.48$\pm$0.01&3.68$\pm$0.02&3.02$\pm$0.02&16.84$\pm$0.29&21.67$\pm$0.33&18.97$\pm$0.28
 \\
CRP  & 5  & 3.08$\pm$0.05&3.57$\pm$0.02&4.29$\pm$0.03&\textbf{0.88$\pm$0.01}&3.47$\pm$0.01&4.03$\pm$0.01&3.18$\pm$0.02&\textbf{19.31$\pm$0.3}&28.3$\pm$0.47&\textbf{20.57$\pm$0.26}
\\
Ours& 5  & \textbf{3.12$\pm$0.05}&\textbf{3.58$\pm$0.02}&\textbf{4.34$\pm$0.02}&\textbf{0.88$\pm$0.01}&\textbf{3.49$\pm$0.01}&\textbf{4.05$\pm$0.01}&\textbf{3.21$\pm$0.01}&18.95$\pm$0.23&\textbf{32.2$\pm$0.43}&19.4$\pm$0.23
 \\
\hlineB{2}
\end{tabularx}
\normalsize
\end{table*}

\begin{table}[!t]
\caption{Performances of the FlowSE with the NFE of 5 trained with flow matching and score matching on WSJ0-CHiME3 dataset. }
\centering
\begin{tabular}{l c c c c c}
\hlineB{2}
METHOD&PESQ&DNSMOS&WVMOS&ESTOI&SI-SDR\\
\hlineB{2}
Flow matching&3.03&4.04&4.00&0.93&18.95\\
Score matching&3.01&4.03& 3.92&0.93&18.57 \\
\hlineB{2}
\label{table:sdevsode}
\end{tabular}
\normalsize 
\end{table}
\section{Experimental Settings}
We use two different datasets, WSJ0-CHiME3 and VoiceBank-DEMAND (VB-DMD), to demonstrate the performance of FlowSE. The WSJ0-CHiME3 dataset was constructed by mixing clean speech utterances from the Wall Street Journal (WSJ0) dataset \cite{garofolo1993csriwsj0} and environmental noises from the CHiME3 dataset \cite{barker2015chime} with the signal-to-noise ratio (SNR) sampled from a uniform distribution between 0 and 20 dB. 
We created the WSJ0-CHiME3 dataset using the source codes\footnote[1]{\texttt{ https://github.com/sp-uhh/sgmse}} provided by the authors of \cite{sgmsep}. 
The dataset consists of a train (12777 files), validation (1206 files) and test sets (615 files). 
The VB-DMD dataset \cite{valentini2016investigatingvctk} is a publicly available dataset generated by mixing clean speech from the VCTK dataset \cite{veaux2013voice} with eight real-recorded noise samples from the DEMAND database \cite{thiemann2013diversedemand} and two artificially generated noise samples (babble and speech shape) at SNRs of 0, 5, 10, and 15 dB. The SNRs for the test set are 2.5, 7.5, 12.5, and 17.5 dB. 
The training dataset is split into a training and validation dataset, using speakers ``p226" and ``p287" for validation, as in \cite{sgmsep, lemercier2023storm, lay2024singlesgmsecrp}.

In this work, the speech signal from the datasets is represented in the complex-valued Short-Time Fourier Transform (STFT) domain after the amplitude compression as in \cite{Lay2023bbed, lay2024singlesgmsecrp}, so $s, y \in\mathbb{C}^{K\times F}$, where $K$ is the number of frames and $F$ is the number of frequency bins.
All configurations of STFT and amplitude compression followed those in \cite{Lay2023bbed, lay2024singlesgmsecrp}.

We adopted the Noise Conditional Score Network (NCSN++) as the architecture of the neural network $v_{\theta}(x_t,y,t) $ with the same configuration as the compared BBED \cite{Lay2023bbed} and CRP \cite{lay2024singlesgmsecrp} models.  $t_\delta$ was set to be $0.03$. 
We trained the neural network $v_{ \theta}(x_t,y,t)$, using Adam optimizer \cite{kingma2014adam} with a learning rate of $10^{-4}$ and a batch size of 8. 
An exponential moving average with a decay of 0.999 was applied to the network parameters. 
$\sigma$ was determined empirically to 0.487 based on the wideband - perceptual evaluation of speech quality (WB-PESQ) \cite{widebandpesq} scores for the validation set. 
We trained the model for 250 epochs and logged the average WB-PESQ scores for 10 random files from the validation set for each epoch to find the best model for speech enhancement. 
The $N$ and $\Delta t(i)$ in the Euler method in (\ref{eq:eulerforSE}) or probability flow ODE in (\ref{eq:pfode})
were set to be $N\in \{1,2,3,4,5\}$, $\Delta t(i)=\frac{t_N-t_\delta}{N-1}$, when $i\in\{1,2,...,N-1\}$ and $\Delta t(N)=t_\delta$  for $N\in \{2,3,4,5\}$. In the case of $N=1$, $\Delta t(1)=t_N$.

We selected two diffusion-based SE models BBED \cite{Lay2023bbed} and CRP \cite{lay2024singlesgmsecrp} as baselines. 
We utilized the pre-trained BBED model available at the model checkpoint\footnote[2]{\texttt{https://github.com/sp-uhh/sgmse\_crp}} shared by the authors of \cite{lay2024singlesgmsecrp} and fine-tuned that model for CRP with $n_{rev} \in \{1,2,3,4,5\}$ with the same configurations in \cite{lay2024singlesgmsecrp}. 

To assess the performance of the proposed method, we have evaluated the WB-PESQ scores\cite{widebandpesq},  Deep Noise Supression MOS (DNSMOS) \cite{salas2013subjective}, wav2vec MOS (WVMOS) \cite{Andreev_2023}, Extended Short-Time Objective Intelligibility (ESTOI) \cite{jensen2016algorithm}, DNSMOS P.835 including SIG, BAK, and OVRL \cite{reddy2022dnsmos}, Scale-Invaraiant Signal-to-Distortion Ratio (SI-SDR), Signal-to-Inference Ratio (SI-SIR), and Signal-to-Artifact Ratio (SI-SAR) \cite{le2019sdr}.

\section{Results}
Table \ref{maintable} summarizes the performances and NFEs for the proposed and compared methods on the WSJ0-CHiME3 and VB-DMD datasets, where the CRP was fine-tuned with $n_{rev}=5$.
The mean of each measure is reported, followed by a 95\% confidence interval. 
We can see that the performances of our method with the NFE of 5 were nearly equivalent to BBED with the NFE of 60 and better than BBED with the NFE of 5.
When compared with CRP, our method performed similarly to CRP in all measurements without any fine-tuning procedure.

\begin{figure}[!t]
\label{figure:vs_sgmsecrp}
    \centering
    \includegraphics[width=0.48\textwidth]{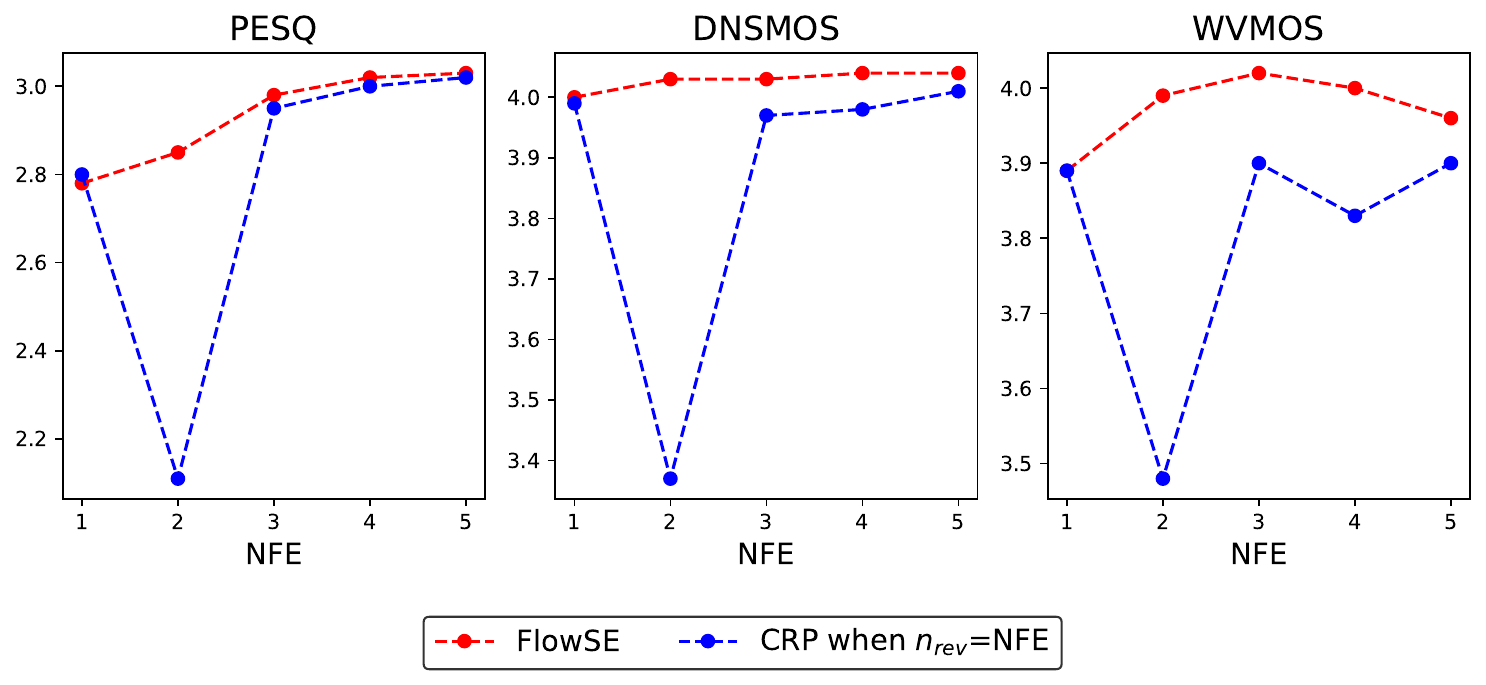} 
    \caption{Performances of the proposed FlowSE and the CRP models fine-tuned with matched $n_{rev}$ for the NFE from 1 to 5 on the WSJ0-CHiME3 dataset. }
    \label{fig:vssgmsecrp}
\end{figure}

For further analysis, we have shown the WB-PESQ scores, DNSMOS, and WVMOS with the NFE from 1 to 5 for the proposed model and the five CRP models fine-tuned with the same $n_{rev}$ with the NFE in the inference phase 
in Fig. \ref{fig:vssgmsecrp}.
Although the proposed FlowSE model was trained without the knowledge of the number of steps in the inference stage, the performances were comparable to the CRP models, each fine-tuned with the matched number of $n_{rev}$. 

Table \ref{table:sdevsode} demonstrates the performances of the proposed FlowSE trained with flow matching and score matching. 
Flow matching indicates that the model $v_\theta(x_t,y,t)$ was trained with the CFM loss in (\ref{eq:cfmenhancement}), while score matching denotes that the score model $s_\theta(x_t,y,t) $ was trained with the DSM loss in $(\ref{eq:dsmloss})$ when the drift and diffusion coefficients were given in \eqref{eq:driftdiffusion}.
It is shown that the performances of two cases with the NFE of 5 were nearly the same. 
It demonstrated that the performance of the BBED with the NFE of 60 could be achieved with the diffusion model with the drift and diffusion coefficients designed from the probability path with linearly moving mean and standard deviation as in the OT conditional vector field, with the NFE of 5 without the need for any fine-tuning procedure. 

\section{Conclusion}
In this work, we propose FlowSE, a speech enhancement model based on conditional flow matching. 
To apply the CFM for SE, we add noisy speech as conditioning variable for the conditional flow and vector field, and the mean and the covariance for the conditional probability path are configured to move linearly from noisy speech to clean speech and decrease linearly, respectively, in a similar way to the OT conditional vector field. 
Experimental results showed that FlowSE with the NFE of 5 achieved performances comparable to the BBED with the NFE of 60 and the CRP using additional fine-tuning procedure with the NFE of 5.
We also derived the diffusion model with the perturbation kernel the same as the conditional probability path for the FlowSE, and demonstrated that the diffusion model with SM achieved similar performances to FlowSE with the NFE of 5 without any fine-tuning procedure. 


\vfill\pagebreak
\bibliographystyle{IEEEtran}
\bibliography{strings,refs}


\end{document}